\documentclass[pra,aps,showkeys,preprint,nofootinbib,superscriptaddress]{revtex4}
\usepackage{graphicx}
\usepackage{amsmath}

\begin{document}
\title{\bf Dynamics of Entanglement and Nonclassical Correlation for Four-Qubit GHZ State}
\author{P. Espoukeh}
\affiliation{Department of Physics, Science and Research Branch,
Islamic Azad University, Tehran, Iran}
\author{R. Rahimi}
\affiliation{Institute for Quantum Computing, University of
Waterloo, Waterloo, ON, N2L 3G1, Canada}
\author{S. Salimi}
\affiliation{Department of Physics, University of Kurdistan, P.O.
Box 66177-15175, Sanandaj, Iran}
\author{P. Pedram}
\affiliation{Department of Physics, Science and Research Branch,
Islamic Azad University, Tehran, Iran}

\begin{abstract}
Many-qubit entanglement is crucial for quantum information
processing although its exploitation is hindered by the detrimental
effects of the environment surrounding the many-qubit system. It is
thus of importance to study the dynamics of general multipartite
nonclassical correlation, including but not restricted to
entanglement, under noise. We did this study for four-qubit GHZ
state under most common noises in an experiment and found that
nonclassical correlation is more robust than entanglement except
when it is imposed to dephasing channel. Quantum discord presents a
sudden transition in its dynamics for Pauli-X and Pauli-Y noises as
well as Bell-diagonal states interacting with dephasing reservoirs
and it decays monotonically for Pauli-Z and isotropic noises.
\end{abstract}

\maketitle

\section{Introduction}
It is assumed that entanglement can be useful in quantum computing
\cite{1}, quantum cryptography \cite{4} and  quantum information
processing such as superdense coding \cite{6} and quantum
teleportation \cite{8}. In practical applications, there are quantum
tasks with no entanglement but representing quantum advantages over
classical counterparts \cite{10,11,13}. This indicates that
entanglement is not the only feature of quantum correlation and we
need to study correlations in a more general concept. This statement
is further more correct once it is an experimental implementation of
quantum computing or quantum information processing. In any
experiment in the field, entanglement is generally difficult to be
achieved, but nonclassical correlation is comparably less
challenging. Due to this fact, and considering the additional fact
that entanglement in larger Hilbert spaces is not well-known yet,
understanding the dynamics of nonclassical correlation, generally,
and entanglement, specifically, is of importance specially from
practical point of view.

Experiments are nowadays elaborated such that  they can, at least in
some physical systems such as NMR \cite{15} and EPR/ENDOR
\cite{16}, achieve controls on larger numbers of quits. Therefore,
it is getting of practical importance to study the characteristics
and behaviour of larger networks of quits and determine the
corresponding dynamics of the correlated states under noises that
are more common in the physical systems of the study.

Quantum discord is one of the measures of nonclassical correlations
\cite{17}. Its definition is based on the difference between two
classically equivalent forms of mutual information. It has been
shown that quantum discord is very useful to describe correlations
involved in quantum systems \cite{19}. Nonclassical correlation and
entanglement can have in principal different behaviours. For
example, in some temperature ranges, quantum discord of a two-qubit
one-dimensional XYZ Heisenberg chain in thermal equilibrium
increases with temperature while entanglement decreases with
temperature and even it goes to zero \cite{20}. Datta et al.
\cite{13} showed that entanglement in DQC1 is negligibly small,
however DQC1 includes nonzero discord. Thus, the notion of quantum
speed up is related to nonclassical correlation rather than to
entanglement \cite{21}.

Implementations of quantum computing and quantum  information
processing are ultimate goals for studies and researches in the
field thus there have been continuously extensive efforts in the
context of performing experiments relating to quantum technologies.
One immediate consideration that is imposed right after starting
plans for experimental practices is how to deal with decoherence of
quantum systems. Decoherence is generally critical for any physical
system that is determined for experiments, but in the context of
quantum protocols this issue becomes one of the most challenging
steps. Decoherence of quantum systems are due to unavoidable
interactions with the environment and it is led to the degradation
of quantum correlations and limitations on using nonclassical
correlation and more importantly quantum entanglement in practical
applications.

Studies on quantum channels are usually divided  into two
categories: Markovian \cite{24,26} and non-Markovian
\cite{19,27,27-5}. Investigations of nonclassical correlations under
Markovian environments show that nonclassical correlation is more
robust than entanglement. Consequently, implementations of quantum
algorithms that rely on nonclassical correlation, and not an
absolute entanglement, are less fragile \cite{28}. Studies on
non-Markovian errors show that since there is no occurrence of
sudden death of quantum discord in spite of the entanglement sudden
death, the description is more practical in the case of nonclassical
correlation \cite{19}.

Time evolutions of nonclassical correlation and  entanglement in
bipartite systems coupled to external environments have been
extensively studied \cite{32}. However, this concept is yet not very
well resolved for higher dimensions. Quantifications of nonclassical
correlation and entanglement for multi-partite systems are not yet
generally d known, in addition and consequently determinations of
their behavior in presence of environmental noises are far from
being clear. In the case of tripartite systems there have been some
results \cite{35,38-1} but for larger spaces the sources are very
few \cite{39,40,40-1} and the existing papers only study the
entanglement of the states not nonclassical correlations in terms of
e.g. quantum discord and their behavior under noise. From the
previous studies, we have learned that nonclassical correlation and
entanglement are preserved in three-qubit GHZ state under noise, but
correlations are eliminated in the case of a W-state \cite{38}.

In this paper, we study nonclassical correlation  and entanglement
of four-qubit Greenberger-Horne-Zeilinger (GHZ) state that is
initially prepared in a pure state described as
$|\mathrm{GHZ}_4\rangle=\frac{1}{\sqrt{2}}\left(|0000\rangle+|1111\rangle\right)$.
The time evolutions are studied for the $|\mathrm{GHZ}_4\rangle$
state under experimentally common noises, Pauli-X (bit-flip),
Pauli-Y (bit-phase-flip), Pauli-Z (phase-flip) and isotropic
(depolarizing) noises. For each type of noise, the time evolution of
the state is given by the solution of the master equation in the
Lindblad form. The entanglement evolution of the mixed state is
described by using the lower bound for multi-qubit concurrence
proposed by Li et al. \cite{41}. Nonclassical correlation of the
mixed four-qubit GHZ state is described by the global quantum
discord that is introduced by Rulli \emph{et al.}  \cite{42}
which uses a systematic extension of bipartite quantum discord.

This paper is organized as follows. In the next  section, we briefly
explain the measures that we use for quantifying entanglement and
nonclassical correlation of  $|\mathrm{GHZ}_4\rangle$ state.  In
Sec.~\ref{sec3}, we obtain time evolutions of entanglement and
nonclassical correlation for the initially prepared
$|\mathrm{GHZ}_4\rangle$ state but imposed to Pauli and isotropic
noises. In Sec.~\ref{sec4}, we present the conclusions.

\section{Entanglement and nonclassical correlation of $|\mathrm{GHZ}_4\rangle$ State}\label{sec2}
There is not any general approach in quantifying  entanglement for
multi-partite states. However, as a practical method, one can make
use of entanglement measures for bipartite states, in which
characterizing entanglement is simpler, and extend them to all
states by convex roof \cite{43}. This approach is generally
difficult to be performed analytically. Here, in order to
characterize entanglement of $|\mathrm{GHZ}_4\rangle$ state, we
apply the lower bound for the multi-qubit concurrence proposed by Li
et al. \cite{41} which is based on the bipartite concurrences of
the multi-qubit system corresponding to all possible bipartite cuts
of the $N$-qubit system.

For a pure $N$-qubit state, the concurrence is defined as follows
\begin{equation}
C_N(|\psi\rangle ) =\sqrt{1-\frac{1}{N}\sum_{i=1}^N \mbox{Tr} \rho_i^ 2}
\end{equation}
where $\rho_i=\mbox{Tr}(|\psi\rangle\langle\psi|)$  is the reduced
density matrix of the $i$-th qubit after tracing out other $N-1$
qubits. The concurrence for mixed $N$-qubit states can be
generalized as follows
\begin{equation}
C_N(\rho ) =\mbox{min} \sum_i p_i C_N(|\psi_i\rangle ),
\end{equation}
in which the minimum is over the pure-state ensemble  that specifies
$\rho$, namely, $\rho =\sum_i p_i |\psi_i\rangle\langle\psi_i| $.
Note, that there is no analytical solution to optimize the
concurrence of a multi-qubit system except for two-qubit systems and
a special case of three-qubit state. Recently, Li et al. \cite{41}
presented a lower bound to this measure which is simpler to
calculate
\begin{equation}\label{lowerbound}
C_N(\rho) \geq \tau_N(\rho)\equiv\sqrt{\frac{1}{N}\sum_{n=1}^N\sum_{k=1}^K (C_k^n)^2}.
\end{equation}
This bound contains $N$ bipartite concurrences $C^n$ that
correspond to the possible bipartite cuts of the multi-qubit system.
In order to calculate each bipartite concurrence $C^n$, one needs to
evaluate a sum of $K= 2^{N-2} (2^{N-1} - 1)$ terms $C_k$ that are
given by
\begin{equation}
C^n_k =\mathrm{max}\{0, \lambda_k^1 - \lambda_k^2 - \lambda_k^3 -\lambda_k^4\}.
\end{equation}
Here, $\lambda_k^i$'s ($i = 1..4$) are the square roots  of the four
non-vanishing eigenvalues of the matrix $\rho \tilde\rho_k^n$ in
decreasing order and $\tilde\rho_k^n=S_k^n\rho^*S_k^n$ in which
$\rho^*$ is the complex conjugate of $\rho$ and $S_k^n =L_k^n
\otimes L_0, k = 1, ...,K$. Moreover, $L_0$ is the generator of the
group $SO(2)$ and $L_k^n$'s are the generators of the group
$SO(2^{N-1})$.

Consider the density matrix of a composed system $AB$ being  denoted
by $\rho$. $\rho^A=\mathrm{Tr}_B(\rho)$ and
$\rho^B=\mathrm{Tr}_A(\rho)$ are the reduced density matrices for
subsystems $A$ and $B$, respectively. Quantum discord for this
system, according to the definition of Ollivier and Zurek, is given
by
\begin{equation}
D(\rho)=I(\rho)- C(\rho),
\end{equation}
where $I(\rho)=S(\rho^A) + S(\rho^B) - S(\rho)$ is the quantum
mutual information that is a well-known measure of the total
correlation and $S(\rho)=-\mathrm{Tr}(\rho \log_{2}\rho)$ is the
von-Neumann entropy. In addition,
\begin{equation}
C(\rho)=S(\rho^a)- \mathrm{min}_{\{\Pi_k\}}S(\rho|\{\Pi_k\})
\end{equation}
is a measure of classical correlations in which the conditional
entropy  is defined as $S(\rho|\{\Pi_k\})=\sum_kp_kS(\rho_k)$ where
$\rho_k=\frac{1}{p_k}(I^a\otimes\Pi_k^b)\rho(I^a\otimes\Pi_k^b)$ is
the conditional density operator,
$p_k=\mathrm{Tr}[(I^A\otimes\Pi_k^B)\rho]$, and the minimum is taken
over the set of projective measurements $\{\Pi_k\}$. This form of
quantum discord is appropriate for bipartite systems and for
multi-qubit states we resort to a proper generalization of it.

There are various approaches to generalize quantum discord for
multipartite  states \cite{44,46,47}. Here, in order to evaluate
quantum discord for a four-qubit system, we apply the global quantum
discord which is introduced by Rulli \emph{et al.}. It uses a
systematic extension of bipartite quantum discord \cite{42}. The
global quantum discord $D(\rho_{A_1...A_N})$ for a multipartite
state $\rho_{A_1...A_N}$ is then defined as
\begin{eqnarray}\label{gqd}
D(\rho_{A_1...A_N})= \mathrm{min}_{\{\Pi_k\}}[S(\rho_{A_1...A_N}\|
\Phi(\rho_{A_1...A_N})) -\sum_{j=1}^N
S(\rho_{A_j}\|\Phi_j(\rho_{Aj}))],
\end{eqnarray}
where $S(\rho_{A_1...A_N}\| \Phi(\rho_{A_1...A_N}))$ is the relative entropy and
\begin{eqnarray}
\Phi_j(\rho_{A_j})&=&\sum_{i}\Pi_{A_j}^{i}\rho_{A_j}\Pi_{A_j}^{i},\\
\Phi(\rho_{A_1...A_N})&=&\sum_k\Pi_k\rho_{A_1...A_N}\Pi_k,\\
\Pi_k&=&\Pi_{A_1}^{j_1}\otimes...\otimes\Pi_{A_N}^{j_N} ,
\end{eqnarray}
and $k$ denotes the index string $(j_1...j_N)$.

In order to define local projective measurements, we select a set of
von-Neumann measurements as
\begin{eqnarray}
\Pi_{A_j}^1={\left(
\begin{matrix}
\cos^2( \frac{\theta_j}{2} ) & e^{i\phi_j} \cos( \frac{\theta_j}{2} ) \sin( \frac{\theta_j}{2} )\\
e^{-i\phi_j} \cos( \frac{\theta_j}{2} ) \sin( \frac{\theta_j}{2} ) & \sin^2( \frac{\theta_j}{2} )
\end{matrix}\right)}
\end{eqnarray}
and
\begin{eqnarray}
\Pi_{A_j}^2=\left(
\begin{matrix}\sin^2( \frac{\theta_j}{2} ) & -e^{-i\phi_j} \cos( \frac{\theta_j}{2} ) \sin( \frac{\theta_j}{2} )\\
e^{i\phi_j} \cos( \frac{\theta_j}{2} ) \sin( \frac{\theta_j}{2} ) & \cos^2( \frac{\theta_j}{2} )
\end{matrix}\right)
\end{eqnarray}
where $\theta_j\in[0,\pi)$  and $\phi_j\in[0,2\pi)$. By varying the
angles $\theta_j$  and $\phi_j$, for $j=1..4$, one can find the
measurement basis that minimizes Eq.~(\ref{gqd}).

\section{Time Evolution of $|\mathrm{GHZ}_4\rangle$ State Under Noise}\label{sec3}
Time evolution of a quantum system under noise is given by the
master equation in the  Lindblad form  \cite{48}
\begin{equation}\label{Lindblad}
\frac{\partial \rho}{\partial t} = -\frac{i}{\hbar} [H_S, \rho] +\sum_{i, \alpha} \left( \mathcal{L}_{i,\alpha} \rho  \mathcal{L}_{i,\alpha}^{\dagger} -
\frac{1}{2} \left\{ \mathcal{L}_{i,\alpha}^{\dagger} \mathcal{L}_{i,\alpha}, \rho\right\} \right),
\end{equation}
where $\mathcal{L}_{i,\alpha}= \sqrt{\kappa_{i,\alpha}}
\sigma^{(i)}_{\alpha}$ is the Lindblad operator that describes the
noise and acts on the $i$th qubit. $\kappa_{i,\alpha}$ is the
decoherence rate and $\sigma^{(i)}_{\alpha}$ are the Pauli spin matrices for the $i$th
qubit with $\alpha=x$ for Pauli-X, $\alpha = y$ for Pauli-Y and $\alpha = z$ for Pauli-Z noise. Also, $H_S$ is the Hamiltonian of the system.
Solutions to the above equation under various noises for four-qubit
GHZ state is recently presented in Ref.~\cite{49}. The idea is to
write the density matrix for infinitesimal time interval $t=\delta
t$ by using the Lindblad equation as
\begin{eqnarray}\label{del}
\rho(\delta t)&=& \rho(0)+\left[ \sum_{i, \alpha} \left(\mathcal{L}_{i,\alpha}
\rho(0) \mathcal{L}_{i,\alpha}^{\dagger} \right) - \frac{1}{2}
\left\{\mathcal{L}_{i,\alpha}^{\dagger} \mathcal{L}_{i,\alpha}, \rho(0) \right\}
\right]\delta t,
\end{eqnarray}
where $\rho(0)= |\mbox{GHZ}_4\rangle\langle\mbox{GHZ}_4|$. Then, by
using  a proper ansatz, the solutions can be obtained for all time
\cite{49}.

Under different noises, if each qubit interacts locally with the
environment  then the dynamics of nonclassical correlation and
entanglement can be extracted from the above formulation,
specifically for an initially prepared four quit GHZ state.

\subsection{Pauli-X noise}
\begin{figure}
\centering
\includegraphics[width=10cm]{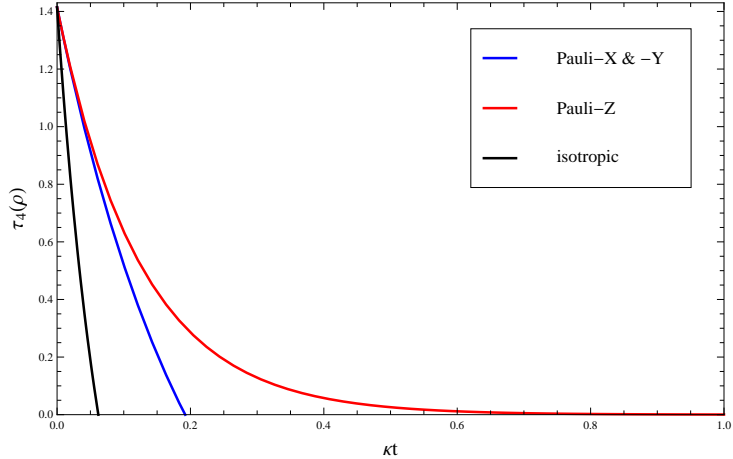}
\caption{\label{fig1} The lower bound for the four-qubit concurrence for an initial $|\mathrm{GHZ}_4\rangle$ state, transmitted through Pauli-X, Pauli-Y, Pauli-Z and isotropic channels as function of $\kappa t$.}
\end{figure}

The density matrix of four quit GHZ state under the Pauli-X error is the solution of the Lindblad equation \cite{49}, as follows
\begin{eqnarray}\label{x}
\rho^{\rm X}= { \left(
\begin{smallmatrix}
\alpha & 0 & 0 & 0 & 0 & 0 & 0 & 0 & 0 & 0 & 0 & 0 & 0 & 0 & 0 & \alpha  \\
0 & \beta & 0 & 0 & 0 & 0 & 0 & 0 & 0 & 0 & 0 & 0 & 0 & 0 & \beta & 0   \\
0 & 0 & \beta & 0 & 0 & 0 & 0 & 0 & 0 & 0 & 0 & 0 & 0 & \beta & 0 & 0   \\
0 & 0 & 0 & \gamma & 0 & 0 & 0 & 0 & 0 & 0 & 0 & 0 & \gamma & 0 & 0 & 0   \\
0 & 0 & 0 & 0 & \beta & 0 & 0 & 0 & 0 & 0 & 0 & \beta & 0 & 0 & 0 & 0   \\
0 & 0 & 0 & 0 & 0 & \gamma & 0 & 0 & 0 & 0 & \gamma & 0 & 0 & 0 & 0 & 0   \\
0 & 0 & 0 & 0 & 0 & 0 & \gamma & 0 & 0 & \gamma & 0 & 0 & 0 & 0 & 0 & 0   \\
0 & 0 & 0 & 0 & 0 & 0 & 0 & \beta & \beta & 0 & 0 & 0 & 0 & 0 & 0 & 0   \\
0 & 0 & 0 & 0 & 0 & 0 & 0 & \beta & \beta & 0 & 0 & 0 & 0 & 0 & 0 & 0   \\
0 & 0 & 0 & 0 & 0 & 0 & \gamma & 0 & 0 & \gamma & 0 & 0 & 0 & 0 & 0 & 0   \\
0 & 0 & 0 & 0 & 0 & \gamma & 0 & 0 & 0 & 0 & \gamma & 0 & 0 & 0 & 0 & 0   \\
0 & 0 & 0 & 0 & \beta & 0 & 0 & 0 & 0 & 0 & 0 & \beta & 0 & 0 & 0 & 0   \\
0 & 0 & 0 & \gamma & 0 & 0 & 0 & 0 & 0 & 0 & 0 & 0 & \gamma & 0 & 0 & 0   \\
0 & 0 & \beta & 0 & 0 & 0 & 0 & 0 & 0 & 0 & 0 & 0 & 0 & \beta & 0 & 0   \\
0 & \beta & 0 & 0 & 0 & 0 & 0 & 0 & 0 & 0 & 0 & 0 & 0 & 0 & \beta & 0   \\
\alpha & 0 & 0 & 0 & 0 & 0 & 0 & 0 & 0 & 0 & 0 & 0 & 0 & 0 & 0 & \alpha
\end{smallmatrix} \right),}
\end{eqnarray}
where
\begin{eqnarray} \left\{
\begin{array}{l}
\alpha =\frac{1}{16} \left(1 + 6 e^{-4 \kappa t} + e^{-8 \kappa t}\right), \\
\beta =\frac{1}{16}\left(1 - e^{-8 \kappa t}\right),\\
\gamma=\frac{1}{16} \left(1 - 2 e^{-4 \kappa t} + e^{-8 \kappa t}\right),
\end{array} \right.
\end{eqnarray}
and $\kappa$ is the decoherence parameter. For this mixed state, the
lower bound Eq.~(\ref{lowerbound}) to the four-qubit concurrence
gives
\begin{equation}\label{entx}
\tau(\rho^{\rm X}) =  \mathrm{max} \left\{0, \frac{\sqrt{2}}{4}\left(e^{-8 \kappa t} + 6 e^{-4 \kappa t} - 3\right)\right\},
\end{equation}
which is an exponentially decaying function of time. The plot of
this lower bound (Fig.~\ref{fig1}) shows that the entanglement of
the  four-qubit GHZ state vanishes after some finite time for the
bit-flip channel. However, using the positive partial transpose
criteria \cite{43}, we find it out that the density matrix
of (\ref{x}) is separable only for $t\rightarrow\infty$. This
indicates that the introduced lower bound does not display proper
long-time entanglement for this state.

In order to obtain nonclassical correlation, we need to evaluate
Eq.~(\ref{gqd}) by first finding the reduced density matrix  of
$i$th subsystem which is obtained by tracing out other three qubits
that reads
\begin{equation}
\rho^{\rm X}_{A_1}=\rho^{\rm X}_{A_2}=\rho^{\rm X}_{A_3}=\rho^{\rm X}_{A_4}=\left(\alpha + 4 \beta + 3 \gamma\right)\textbf{1}.
\end{equation}
Since $\rho^{\rm X}_{A_j}$ is proportional to the unity operator,
$\Phi_j(\rho_{A_j})=\sum_{i}\Pi_{A_j}^{i}\rho_{A_j}\Pi_{A_j}^{i}$ is
unaffected under any projective measurements, therefore,
$\Phi_j(\rho^{\rm X}_{A_j})=\rho^{\rm X}_{A_j}$ and the second term
in Eq.~(\ref{gqd}) vanishes. To evaluate the first term, we need to
find the measurement bases that minimize quantum discord.
Calculations show that quantum discord suddenly changes during the
dynamics of the system, namely, the measurement bases change from
$\sigma_z$ ($\theta=0$) to $\sigma_x$ ($\theta=\pi/2$), so we find
\begin{eqnarray}
S(\Phi({\rho^{\rm X}}))&=& 4-\frac{(1 - e^{-8 \kappa t})}{2}\log_2
(1 - e^{-8 \kappa t})\nonumber \\
&&-\frac{1}{8}[ (1+ 6 e^{-4 \kappa t}+ e^{-8 \kappa t})\log_2 (1+ 6 e^{-4 \kappa t}+ e^{-8 \kappa t})\nonumber \\
&&+ 3(1- 2 e^{-4 \kappa t}+ e^{-8 \kappa t})\log_2 (1- 2 e^{-4\kappa
t}+ e^{-8 \kappa t})].
\end{eqnarray}
for $0<\kappa t<0.137$ and
\begin{equation}
S(\Phi({\rho^{\rm X}})) = 3,
\end{equation}
for $\kappa t>0.137$. Also,
\begin{eqnarray}
S({\rho^{\rm X}})&=&3 -\frac{(1 - e^{-8 \kappa t})}{2}\log_2 (1 -
e^{-8 \kappa t})\nonumber \\
&&-\frac{1}{8}[ (1+ 6 e^{-4 \kappa t}+ e^{-8 \kappa t})\log_2 (1+ 6 e^{-4 \kappa t}+ e^{-8 \kappa t})\nonumber \\
&&+ 3(1- 2 e^{-4 \kappa t}+ e^{-8 \kappa t})\log_2 (1- 2 e^{-4
\kappa t}+ e^{-8 \kappa t})].
\end{eqnarray}
Finally, we obtain
\begin{equation}\label{dx1}
D({\rho^{\rm X}})=1,
\end{equation}
for $0<\kappa t<0.137$ and
\begin{eqnarray}\label{dx2}
D({\rho^{\rm X}})&=& \frac{(1 - e^{-8 \kappa t})}{2}\log_2 (1 -
e^{-8 \kappa t})\nonumber \\
&&+\frac{1}{8}[ (1+ 6 e^{-4 \kappa t}+ e^{-8 \kappa t})\log_2 (1+ 6 e^{-4 \kappa t}+ e^{-8 \kappa t})\nonumber \\
&&+ 3(1- 2 e^{-4 \kappa t}+ e^{-8 \kappa t})\log_2 (1- 2 e^{-4
\kappa t}+ e^{-8 \kappa t})],
\end{eqnarray}
for $\kappa t>0.137$. Therefore, the quantum discord is constant for
$0<\kappa t<0.137$, however, it experiences a sudden transition in
its dynamics at $\kappa t=0.137$ and monotonically decays for
$\kappa t>0.137$. This peculiar behavior of quantum discord is known
as the sudden change of the quantum discord which is also observed
for two qubit systems \cite{51,53-2}, linked to the universal
freezing of quantum correlations \cite{53-5}.

\subsection{Pauli-Y noise}
Now, we investigate the time evolution of nonclassical correlation
and entanglement for $|\mathrm{GHZ}_4\rangle$  state under Pauli-Y
error. The density matrix for this system is given by \cite{49}
\begin{eqnarray}
{\rho^{\rm Y}}= { \left(
\begin{smallmatrix}
\alpha & 0 & 0 & 0 & 0 & 0 & 0 & 0 & 0 & 0 & 0 & 0 & 0 & 0 & 0 & \alpha  \\
0 & \beta & 0 & 0 & 0 & 0 & 0 & 0 & 0 & 0 & 0 & 0 & 0 & 0 & -\beta & 0   \\
0 & 0 & \beta & 0 & 0 & 0 & 0 & 0 & 0 & 0 & 0 & 0 & 0 &- \beta & 0 & 0   \\
0 & 0 & 0 & \gamma & 0 & 0 & 0 & 0 & 0 & 0 & 0 & 0 &- \gamma & 0 & 0 & 0   \\
0 & 0 & 0 & 0 & \beta & 0 & 0 & 0 & 0 & 0 & 0 & -\beta & 0 & 0 & 0 & 0   \\
0 & 0 & 0 & 0 & 0 & \gamma & 0 & 0 & 0 & 0 & \gamma & 0 & 0 & 0 & 0 & 0   \\
0 & 0 & 0 & 0 & 0 & 0 & \gamma & 0 & 0 & \gamma & 0 & 0 & 0 & 0 & 0 & 0   \\
0 & 0 & 0 & 0 & 0 & 0 & 0 & \beta & -\beta & 0 & 0 & 0 & 0 & 0 & 0 & 0   \\
0 & 0 & 0 & 0 & 0 & 0 & 0 &- \beta & \beta & 0 & 0 & 0 & 0 & 0 & 0 & 0   \\
0 & 0 & 0 & 0 & 0 & 0 & \gamma & 0 & 0 & \gamma & 0 & 0 & 0 & 0 & 0 & 0   \\
0 & 0 & 0 & 0 & 0 & \gamma & 0 & 0 & 0 & 0 & \gamma & 0 & 0 & 0 & 0 & 0   \\
0 & 0 & 0 & 0 & -\beta & 0 & 0 & 0 & 0 & 0 & 0 & \beta & 0 & 0 & 0 & 0   \\
0 & 0 & 0 & \gamma & 0 & 0 & 0 & 0 & 0 & 0 & 0 & 0 & \gamma & 0 & 0 & 0   \\
0 & 0 &- \beta & 0 & 0 & 0 & 0 & 0 & 0 & 0 & 0 & 0 & 0 & \beta & 0 & 0   \\
0 & -\beta & 0 & 0 & 0 & 0 & 0 & 0 & 0 & 0 & 0 & 0 & 0 & 0 & \beta & 0   \\
\alpha & 0 & 0 & 0 & 0 & 0 & 0 & 0 & 0 & 0 & 0 & 0 & 0 & 0 & 0 & \alpha
\end{smallmatrix} \right).}
\end{eqnarray}
For this case, the lower bound to the concurrence is similar to
Eq.~(\ref{entx}), i.e., $\tau({\rho^{\rm Y}}) =  \mathrm{max}
\left\{0, \frac{\sqrt{2}}{4} \left(e^{-8 \kappa t} + 6 e^{-4 \kappa
t} - 3\right)\right\}$. Also, similar to Pauli-X noise, we observe a
sudden change in quantum discord at $\kappa t=0.137$. Indeed, for
$\kappa t<0.137$ the optimal local measurements are given by
$\Pi_{A_j}^i=| i\rangle\langle i|$ and since $\Phi_j(\rho^{\rm
Y}_{A_j})=\rho^{\rm Y}_{A_j}$ and $S(\Phi({\rho^{\rm
Y}}))-S({\rho^{\rm Y}})=1$, we find
\begin{equation}\label{dy}
D({\rho^{\rm Y}})=1.
\end{equation}
But, for $\kappa t>0.137$, $\sigma_y$ is the basis which minimizes
quantum discord, so we have $S(\Phi({\rho^{\rm Y}}))=3$ and
$S({\rho^{\rm Y}})=S({\rho^{\rm X}})$. Therefore, $D({\rho^{\rm
Y}})$ coincides with Eq.~(\ref{dx2}).

\subsection{Pauli-Z noise}
For the next case, consider Pauli-Z error where its corresponding
density matrix takes the following form \cite{49}
\begin{eqnarray}\label{z}
{\rho^{\rm Z}} = \frac{1}{2} \left(|0\rangle^{\otimes 4} \langle 0
|^{\otimes 4} + |1\rangle ^{\otimes 4}\langle 1|^{\otimes 4}\right)
 + \frac{1}{2}  e^{-8\kappa t} \left(|0\rangle ^{\otimes 4} \langle 1| ^{\otimes 4}+ |1\rangle ^{\otimes 4}\langle 0 |^{\otimes 4} \right),
\end{eqnarray}
that results in the lower bound
\begin{equation}\label{entz}
\tau({\rho^{\rm Z}}) = \sqrt{2}\, e^{-8 \kappa t}.
\end{equation}
As it is shown in Fig.~\ref{fig1}, nonclassical correlation and
entanglement of this state decrease more slowly with respect  to the
previous noise models. For this case, the reduced density matrices
become
\begin{equation}
\rho^{\rm Z}_{A_1}=\rho^{\rm Z}_{A_2}=\rho^{\rm Z}_{A_3}=\rho^{\rm Z}_{A_4}=\frac{\textbf{1}}{2},
\end{equation}
which results in $\Phi_j(\rho^{\rm Z}_{A_j})=\rho_{A_j}$, therefore
the second term of Eq.~(\ref{gqd}) vanishes. The  optimization
procedure of $\theta$ and $\phi$ for each qubit shows that the
optimized projective measurements are the bases of $\sigma_z$. Then
we find $S(\Phi({\rho^{\rm Z}}))=1$ and the von-Neumann entropy
becomes
\begin{eqnarray}
S({\rho^{\rm Z}})= 1-\frac{1}{2}(1 - e^{-8 \kappa t})\log_2 (1-
e^{-8 \kappa t}) +\frac{1}{2}(1 + e^{-8 \kappa t})\log_2 (1 + e^{-8
\kappa t}).
\end{eqnarray}
Therefore, we have
\begin{eqnarray}\label{dz}
D({\rho^{\rm Z}})=\frac{1}{2}(1 - e^{-8 \kappa t})\log_2  (1 - e^{-8
\kappa t}) +(1 + e^{-8 \kappa t})\log_2 (1 + e^{-8 \kappa t}).
\end{eqnarray}
This result shows that the Pauli-Z error causes dissipative behavior
on nonclassical correlation  for $|\mathrm{GHZ}_4\rangle$ state.
Decoherence behavior in terms of $\kappa t$ is depicted in
Fig.~\ref{fig2}.

\subsection{Isotropic noise}
To this end, consider $|\mathrm{GHZ}_4\rangle$ state under isotropic
noise. It should be noted that for this case, the Lindblad equation depends on all three Pauli operators (see Ref. 29). Thus, the density matrix as the solution of the master
equation is described by \cite{49}
\begin{eqnarray}\label{rod}
{\rho^{\rm D}}= { \left(
\begin{smallmatrix}
\tilde\alpha_+ & 0 & 0 & 0 & 0 & 0 & 0 & 0 & 0 & 0 & 0 & 0 & 0 & 0 & 0 &\tilde \alpha_-  \\
0 & \tilde\beta & 0 & 0 & 0 & 0 & 0 & 0 & 0 & 0 & 0 & 0 & 0 & 0 & 0 & 0   \\
0 & 0 & \tilde\beta & 0 & 0 & 0 & 0 & 0 & 0 & 0 & 0 & 0 & 0 &0 & 0 & 0   \\
0 & 0 & 0 & \tilde\gamma & 0 & 0 & 0 & 0 & 0 & 0 & 0 & 0 &0 & 0 & 0 & 0   \\
0 & 0 & 0 & 0 & \tilde\beta & 0 & 0 & 0 & 0 & 0 & 0 & 0 & 0 & 0 & 0 & 0   \\
0 & 0 & 0 & 0 & 0 & \tilde\gamma & 0 & 0 & 0 & 0 & 0 & 0 & 0 & 0 & 0 & 0   \\
0 & 0 & 0 & 0 & 0 & 0 & \tilde\gamma & 0 & 0 & 0 & 0 & 0 & 0 & 0 & 0 & 0   \\
0 & 0 & 0 & 0 & 0 & 0 & 0 & \tilde\beta & 0 & 0 & 0 & 0 & 0 & 0 & 0 & 0   \\
0 & 0 & 0 & 0 & 0 & 0 & 0 &0 & \tilde\beta & 0 & 0 & 0 & 0 & 0 & 0 & 0   \\
0 & 0 & 0 & 0 & 0 & 0 & 0 & 0 & 0 & \tilde\gamma & 0 & 0 & 0 & 0 & 0 & 0   \\
0 & 0 & 0 & 0 & 0 & 0 & 0 & 0 & 0 & 0 & \tilde\gamma & 0 & 0 & 0 & 0 & 0   \\
0 & 0 & 0 & 0 & 0 & 0 & 0 & 0 & 0 & 0 & 0 & \tilde\beta & 0 & 0 & 0 & 0   \\
0 & 0 & 0 & 0 & 0 & 0 & 0 & 0 & 0 & 0 & 0 & 0 & \tilde\gamma & 0 & 0 & 0   \\
0 & 0 & 0 & 0 & 0 & 0 & 0 & 0 & 0 & 0 & 0 & 0 & 0 & \tilde\beta & 0 & 0   \\
0 & 0 & 0 & 0 & 0 & 0 & 0 & 0 & 0 & 0 & 0 & 0 & 0 & 0 & \tilde\beta & 0   \\
\tilde\alpha_- & 0 & 0 & 0 & 0 & 0 & 0 & 0 & 0 & 0 & 0 & 0 & 0 & 0 & 0 & \tilde\alpha_+
\end{smallmatrix} \right),}
\end{eqnarray}
where
\begin{eqnarray}\left\{
\begin{array}{l}
\tilde\alpha_+ =\frac{1}{16}\Big( 1 + 6 e^{-8 \kappa t} + e^{-16 \kappa t}\Big),\\
\tilde\beta =\frac{1}{16}\Big( 1 - e^{-16 \kappa t}\Big),\\
\tilde\gamma = \frac{1}{16}\Big(1 - 2 e^{-8 \kappa t} +  e^{-16 \kappa t}\Big),\\
\tilde\alpha_- = \frac{1}{2} e^{-16 \kappa t}.
\end{array}\right.
\end{eqnarray}
Now, the lower bound (\ref{lowerbound}) to the concurrence reads
\begin{equation}\label{entd}
\tau_({\rho^{\rm D}}) =  \mathrm{max} \left\{0, \frac{\sqrt{2}}{8}\left(9 e^{-16 \kappa t} + 6 e^{-8 \kappa t} - 7\right)\right\},
\end{equation}
This lower bound shows strong dissipation of entanglement in
presence of isotropic error. It should be noted that for this
density matrix, similar to the Pauli-X and Pauli-Y errors, according
to the positive partial trace criteria, expression (\ref{entd}) is
only valid for $t<t_0$.

\begin{figure}
\centering
\includegraphics[width=10cm]{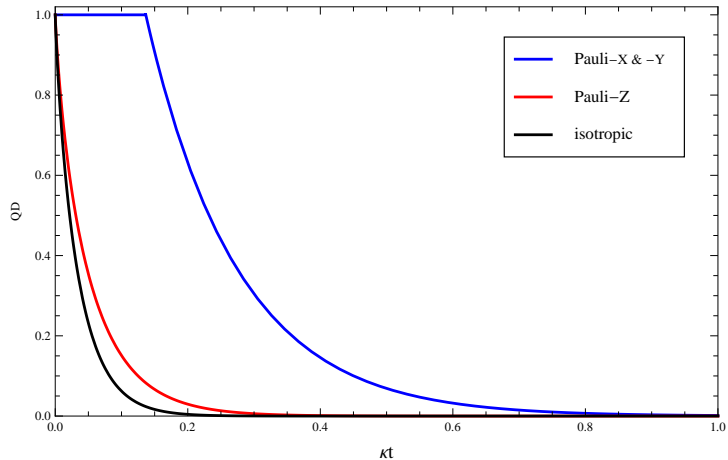}
\caption{\label{fig2} Quantum discord for the four-qubit system with the initial $|\mathrm{GHZ}_4\rangle$ state transmitted under Pauli-X, Pauli-Y, Pauli-Z, and isotropic errors as a function of $\kappa t$.}
\end{figure}

The reduced density matrices become
\begin{equation}
\rho^{\rm D}_{A_1}=\rho^{\rm D}_{A_2}=\rho^{\rm D}_{A_3}=\rho^{\rm D}_{A_4}=\left(\alpha + 4 \beta + 3 \gamma\right)\textbf{1}.
\end{equation}
So, Eq.~(\ref{gqd}) reduces to
\begin{equation}\label{gqdd}
D(\rho^{\rm D}_{A_1...A_N})= \mathrm{min}_{\{\Pi_k\}}S(\rho^{\rm D}_{A_1...A_N}\| \Phi(\rho^{\rm D}_{A_1...A_N})).
\end{equation}
Calculations show that the optimized local projective measurements are $\Pi_{A_j}^i=| i\rangle\langle i|$. Thus, we have
\begin{eqnarray}
S({\rho^{\rm D}})&=& 4-\frac{(1 - e^{-16 \kappa t})}{2}\log_2 (1 -e^{-16 \kappa t})\nonumber \\
&&-\frac{3}{8}(1-2 e^{-8 \kappa t}+ e^{-16 \kappa t})\log_2 (1-2 e^{-8 \kappa t}+ e^{-16 \kappa t})\nonumber \\
&&- \frac{1}{16}(1+6 e^{-8 \kappa t}-7 e^{-16 \kappa t})\log_2 (1+6e^{-8 \kappa t}-7 e^{-16 \kappa t}) \nonumber \\
&& +\frac{1}{16}(1+6 e^{-8 \kappa t}+ 9e^{-16
\kappa t})\log_2 (1+6 e^{-8 \kappa t}+ 9e^{-16 \kappa
t}),\hspace{1cm}\label{sd1}
\end{eqnarray}
and
\begin{eqnarray}
S(\Phi({\rho^{\rm D}}))&=& 4-\frac{(1 - e^{-16 \kappa t})}{2}\log_2(1 - e^{-16 \kappa t})\nonumber \\
&&-\frac{3}{8}(1-2 e^{-8 \kappa t}+ e^{-16 \kappa t})\log_2 (1-2 e^{-8 \kappa t}+ e^{-16 \kappa t})\nonumber \\
&& -\frac{1}{8}(1+6 e^{-8 \kappa t}+ e^{-16 \kappa t})\log_2 (1+6
e^{-8 \kappa t}+ e^{-16 \kappa t}).\label{sd2}
\end{eqnarray}
By inserting Eqs.~(\ref{sd1}) and (\ref{sd2}) in Eq.~(\ref{gqdd}) then
\begin{eqnarray}
D({\rho^{\rm D}})&=& - \frac{1}{8}(1+6 e^{-8 \kappa t}+ e^{-16\kappa t}) \log_2 (1+6 e^{-8 \kappa t}+ e^{-16 \kappa t})\nonumber \\
&&+\frac{1}{16}(1+6 e^{-8 \kappa t}-7 e^{-16 \kappa t})\log_2 (1+6 e^{-8 \kappa t}-7 e^{-16 \kappa t})\nonumber \\
&&+\frac{1}{16}(1+6 e^{-8 \kappa t}+ 9e^{-16 \kappa t})\log_2 (1+6 e^{-8 \kappa
t}+ 9e^{-16 \kappa t})].\label{dd}
\end{eqnarray}
The time evolution of quantum discord for this case is plotted in
Fig.~\ref{fig2}. As the figure shows, the isotropic error results in
a more decrease in quantum discord in comparison to Pauli-Z channel.

\section{Conclusions}\label{sec4}
In this paper, we investigated  the dynamics of nonclassical
correlation and entanglement for the initial
$|\mathrm{GHZ}_4\rangle$  state under interactions with several
independent Markovian environments. These noises include Pauli-X,
Pauli-Y, Pauli-Z and the isotropic errors. In order to describe the
time evolution of entanglement, we utilized the lower bound
(\ref{lowerbound}) as an approximation for the convex roof of the
four-qubit concurrence. The results showed that entanglement is more
robust in transmission under the Pauli-Z error compared to Pauli-X,
Pauli-Y and isotropic errors. For Pauli-X, Pauli-Y, and isotropic
errors the lower bound vanishes in a finite time (sudden death
phenomenon). However, from the positive partial transpose
separability criteria \cite{43}, we conclude that these
states are separable only for $t\rightarrow\infty$. Thus, the lower
bound only describes the behavior of the entanglement for $t<t_0$
where the lower bound vanishes at $t_0$. This is due to the fact
that the accuracy of the lower bound over convex roof seems to
depend on the rank of the corresponding density matrix
\cite{36}. Indeed, the numerical calculation of the convex
roof is a quite difficult task which requires an optimization over
too many (proportional to $r^3$ where $r$ is the rank of the density
matrix) free parameters \cite{36}.

In order to obtain the time evolution of nonclassical correlation,
we used the global quantum discord as a  measure to calculate
nonclassical correlations in multipartite states. We found that
quantum discord is not affected by Pauli-X and Pauli-Y noises for
$\kappa t<0.137$ whereas it decreases monotonically for $\kappa
t>0.137$. This transition of quantum discord is due to the transition
of measurement bases which minimize the quantum discord. Also,
quantum discord decays in the presence of Pauli-Z and isotropic
noises in terms of the dimensionless scaled time $\kappa t$. In
comparison, quantum discord of $|\mathrm{GHZ}_3\rangle$ state is
more robust than $|\mathrm{GHZ}_4\rangle$ against decoherence for
all noises under study \cite{54}. Moreover, except the case of
Pauli-Z error, quantum discord is more robust than entanglement for
the initially prepared $|\mathrm{GHZ}_4\rangle$ state.

\acknowledgements RR is supported by CIFAR, Industry Canada and
NSERC.


\end{document}